\documentclass[aps]{revtex4}
    \def\be{\begin{equation}}
    \def\ee{\end{equation}}
    \def\ba{\begin{eqnarray}}
    \def\ea{\end{eqnarray}}

    \def\e{\mbox{e}}

    \def\alphat{\tilde{\alpha}}
    \def\betat{\tilde{\beta}}

  \input epsf
\usepackage{pstcol,graphicx}
\begin{document}
\title{Second Order Gauge-Invariant Perturbations during Inflation}

\author{F. Finelli $^{1 \,, 2 \,, 3}$} \email{finelli@iasfbo.inaf.it}
\affiliation{$^1$ INAF/IASF-BO, Istituto di Astrofisica Spaziale e 
    Fisica Cosmica di Bologna \\ Istituto Nazionale di Astrofisica \\ 
Via Gobetti 101, I-40129 Bologna, Italy}
\affiliation{$^2$ INAF/OAB, Osservatorio Astronomico di Bologna \\ Istituto 
Nazionale di Astrofisica \\
Via Ranzani 1, I-40127 Bologna, Italy}
\affiliation{$^3$ INFN, Sezione di Bologna, 
Via Irnerio 46, I-40126 Bologna, Italy}

\author{G. Marozzi $^4$} \email{marozzi@bo.infn.it}
\author{G. P. Vacca $^4$} \email{vacca@bo.infn.it}
\author{G. Venturi $^4$} \email{armitage@bo.infn.it}
\affiliation{$^4$ Dipartimento di Fisica, Universit\`a degli Studi di Bologna
    and INFN, \\ via Irnerio 46, I-40126 Bologna, Italy}
\begin{abstract}
The evolution of gauge invariant second-order scalar perturbations
in a general single field inflationary scenario are presented. 
Different second order gauge invariant expressions for the curvature are 
considered.
We evaluate {\em perturbatively} one of these 
second order curvature fluctuations 
and a second order gauge invariant scalar field fluctuation  
during the slow-roll stage of a massive chaotic inflationary scenario, 
taking into account the deviation from a pure de Sitter evolution and 
considering only the contribution of super-Hubble perturbations 
in mode-mode coupling. 
The spectra resulting from their contribution to the
second order quantum correlation function are nearly scale-invariant, 
with additional logarithmic corrections with respect 
to the  first order spectrum. 
For all scales of interest the amplitude of these spectra depends on the total 
number of e-folds. We find, on comparing first and second order perturbation 
results, an upper limit to the total number of e-folds beyond which the 
two orders are comparable. 
\end{abstract}

\maketitle
\section{Introduction}
One of the observational arguments supporting inflation is the 
evidence of nearly scale-invariant primordial adiabatic fluctuations with 
a mostly gaussian spectrum, as shown by COBE \cite{cobe-nongauss} and 
WMAP \cite{wmap-nongauss,wmap06}. In its simplest realizations, adiabatic 
cosmological fluctuations are simply related to linear 
fluctuations of the inflaton generated during the accelerated stage: 
such fluctuations have a gaussian spectrum. 

Non-linear fluctuations inevitably carry a $\chi$-squared distribution, 
which may contribute to non-gaussianities in the 3-D power spectrum and in CMB 
anisotropies depending on the amplitude 
and spectrum of these non-linearities (see \cite{BKMR_report} for a review).
A study of fluctuations beyond the linear level is therefore interesting 
not only for the issue of the stability of inflationary dynamics, 
but also because of its connection with the 
predictions of non-gaussianities in the matter power spectrum and 
in CMB anisotropies.
Even in absence of non-linearity in the potential for the scalar field, 
the generality of intrinsic non-gaussianities is guaranteed by the 
non-linearity of the Einstein equations, a fact which has motivated a great 
interest both from the theoretical and observational point of view
\cite{BKMR_report}.
Our choice of calculating the intrinsic non-gaussianities for a free 
massive inflationary model is therefore a case of primary interest, 
since in such a case the whole non-linear sector is given by gravity.

In this paper we shall focus on the perturbative evualuation of {\em second 
order gauge-invariant} (GI henceforth) curvature fluctuations. 
For this purpose we discuss different 
second order GI measures of curvature perturbations (see also 
\cite{ABMR,NohHwang,MW,LMS,vernizzi,LV_prl,LythRodriguez}), 
which to first order coincide with the curvature on flat 
slicing hypersurfaces. 
We then compute the correlation function for differents 
second order GI variables, finding, for all scales of interest,
a growth in the spectrum amplitude as 
the total number of e-folds which inflation lasted increases. A dependence 
on the total number of e-folds was also found for the production 
of test scalar fields with a mass smaller than that of 
the inflaton \cite{FMVV_I}.

The paper is organized as follows. We present the formalism and 
different gauge-invariant measures of curvature perturbations in section 
II. In sections III and IV we evaluate the quantum correlators for 
different second order GI variables with both analytical and numerical 
methods. We conclude in section V and in appendices A e B we add 
useful technical formulae for gauge transformations and for the analytical 
results given in section III.

\section{General Curvature Perturbation}\label{one}
Let us consider inflation in a flat universe driven by a classical minimally 
coupled scalar field with a potential $V(\phi)$ and follow the notation 
in \cite{FMVV_2}. 
The action is given by:
    \be
    S \equiv \int d^4x {\cal L} = \int d^4x \sqrt{-g} \left[ 
\frac{R}{16{\pi}G}
    - \frac{1}{2} g^{\mu \nu}
    \partial_{\mu} \phi \partial_{\nu} \phi
    - V(\phi) \right] \,
    \label{action}
\ee
where ${\cal L}$ is the Lagrangian density and we shall study the chaotic 
model $V(\phi)=\frac{m^2}{2}\phi^2$ \cite{linde} (see 
\cite{BU} for a second order calculation in a different inflationary
model). We shall consider, up to second order in the 
fluctuations, both the inflaton: 
$$
\phi(t, {\bf x})=\phi(t)+\varphi(t, {\bf x})+\varphi^{(2)}(t, {\bf x})\,
\label{inflaton_field}
$$
and the metric of a flat universe ($ds^2 = g_{\mu \nu} d x^\mu d x^\nu$):
\begin{eqnarray}
g_{00} &=& - 1 - 2 \alpha - 2 \alpha^{(2)} \nonumber \\
g_{0i} &=& - \frac{a}{2} \left( \beta_{,i} + \beta_{,i}^{(2)} \right) 
\nonumber \\
g_{ij} &=& a^2 \left[ \delta_{ij} -2 \delta_{ij}  \left( \psi + \psi^{(2)}
 \right) + D_{i j}\left( E + E^{(2)}\right) \right] \,,
\label{metric_second}
\end{eqnarray}
where $D_{ij}=\partial_i \partial_j -1/3 \, \nabla^2 \delta_{ij}$. 
As is clear from the above we limit ourselves to the scalar sector of metric 
perturbations, neglecting vector and tensor perturbations.

In the above formulation the gauge is not fixed, and 
one can eliminate two scalar functions among the four metric coefficients 
and the inflaton fluctuation. 
We wish to extend to second order the GI potential for the curvature
perturbation. 
For this purpose, let us note that to first order the so called
curvature (potential) perturbation 
is given by 
\be
\hat{\psi}= \psi+\frac{1}{6} \nabla^2 E
\label{CP_order1}
\ee
and this quantity is not GI (for the general gauge trasformation see 
appendix A). 
We now  consider the comoving slicing \cite{riotto}, which is 
defined to be the slicing 
orthogonal to the worldlines of comoving observers: they are free-falling and
the expansion defined by them is isotropic. That is the observers do not 
measure any flux of energy ($T^{0\,(1)}_{\,\,i}=0$) and for the universe under
consideration this corresponds to having $\varphi=0$. The transformation from a
general slicing to a comoving slicing with $\varphi=0$ is:
\be
\tilde{\varphi}_{com}= \varphi - \epsilon^0_{(1)} \dot{\phi}=0 \rightarrow 
\epsilon^0_{(1)}
=\frac{\varphi}{\dot{\phi}} 
\label{Trasf_inf_ord_uno_comoving}
\ee
and one obtains 
\be
R^{(1)}=\tilde{\hat{\psi}}_{com}=\psi+\frac{H}{\dot{\phi}}\varphi
+\frac{1}{6}\nabla^2 E \,,
\label{ComovingRord1}
\ee
which is the first order GI comoving curvature (potential) perturbation.

To second order the situation is more involved due to the presence of terms
quadratic in the first order. 
We are interested in the definition of GI quantities including second 
order. Their construction is not unique and 
we shall therefore construct three different GI second order curvature 
(potential) perturbations.

\subsection{The comoving curvature potential perturbation $R_A$}

This variable is obtained by repeating consistently up to second order
the procedure used for the first order comoving curvature (potential)
perturbation.
Starting from
\be
\hat{\psi}^{(2)}= \psi^{(2)}+\frac{1}{6} \nabla^2 E^{(2)}  \,,
\label{CP_order2_linear}
\ee
to second order the comoving slicing is caracterized by $T^{0\,(1) }_{\,
\,i}=0$ 
and $T^{0\, (2)}_{\,\,i}=0$, that is 
$\varphi=0$ 
and $\varphi^{(2)}=0$. In this case, in order to fix all the degrees
of freedom, we 
restrict ourselves to considering only an infinitesimal time trasformation 
up to second order, which implies taking $\epsilon^s_{(1)}$ 
and $\epsilon^s_{(2)}$  equal to zero.
From the first order we obtain the condition in 
Eq.(\ref{Trasf_inf_ord_uno_comoving}), while from the second order we have
\be
\tilde{\varphi}^{(2)}_{com}= \varphi^{(2)} - \epsilon^0_{(1)} \dot{\varphi} 
+ \frac{1}{2} \left[\epsilon^0_{(1)} \left(\epsilon^0_{(1)}\dot{\phi}\right)^.
- \epsilon^0_{(2)} \dot{\phi}\right]=0 \rightarrow 
\epsilon^0_{(2)}
=\frac{2}{\dot{\phi}}\left[\varphi^{(2)} - \epsilon^0_{(1)} \dot{\varphi} 
+ \frac{1}{2} \epsilon^0_{(1)}
\left(\epsilon^0_{(1)}\dot{\phi}\right)^.\right]\,.
\label{Trasf_inf_ord_due_comoving}
\ee
Thus one obtains 
\begin{eqnarray}
R^{(2)}_A=\tilde{\hat{\psi}}^{(2)}_{com}&=&\psi^{(2)}
+\frac{H}{\dot{\phi}}\varphi^{(2)}+\frac{1}{6}\nabla^2 E^{(2)}-
\frac{\varphi}{\dot{\phi}}\left( \dot{\psi}+2 H \psi \right)-
\frac{H}{\dot{\phi}^2}\varphi \dot{\varphi}+\frac{\varphi^2}{\dot{\phi}^2}
\left( -H^2 -\frac{\dot{H}}{2}+\frac{H}{2}\frac{\ddot{\phi}}{\dot{\phi}}
\right) \nonumber \\
& & + \frac{1}{6 a^2}\frac{1}{\dot{\phi}^2}\left[\partial^i\varphi
\partial_i\varphi-\frac{1}{2}\nabla^2\left(\varphi^2\right)\right]-
\frac{1}{6 a}\frac{1}{\dot{\phi}}\left[\partial^i\beta
\partial_i\varphi-\frac{1}{2}\nabla^2\left(\beta \varphi \right)\right]+
\frac{1}{4}
\frac{\partial^i \partial^j}{\nabla^2}\left[-2 H 
\frac{\varphi}{\dot{\phi}}
 D_{ij} E  \right. \nonumber \\
& & \left.
- \frac{\varphi}{\dot{\phi}} D_{ij} \dot{E}
+ \frac{1}{a^2}  \frac{\varphi}{\dot{\phi}^2}
D_{ij} \varphi -
 \frac{1}{2a} \frac{1}{\dot{\phi}} \left(
\varphi D_{ij} \beta + \beta  D_{ij} \varphi
\right) \right]\,.
\label{ComovingRord2}
\end{eqnarray}
Let us note that this quantity is analogous to the second order curvature
perturbation on a uniform density hypersurface defined in \cite{MW}.
These two quantities agree (apart from a sign) in the long wavelength limit 
(see also \cite{vernizzi,LythRodriguez}).

\subsection{The curvature potential perturbation $R_B$}

We give the second order GI curvature perturbation potential 
obtained from a form analogous to
Eq. (\ref{ComovingRord1}) (the first three terms of the following equation), 
and add a 
minimal set of terms necessary to obtain a second order GI variable. 
Using the general transformations of Appendix A, we obtain
\begin{eqnarray}
R^{(2)}_B &=& 
 \psi^{(2)} + \frac{H}{\dot{\phi}} \varphi^{(2)} +\frac{1}{6} \nabla^2 E^{(2)}
\nonumber \\
& & +\frac{1}{2}\left ( 2 H^2 +\dot{H}-H\frac{\ddot{\phi}}{\dot{\phi}}
\right)^{-1}\left( \dot{\psi}+2 H \psi +\frac{H}{\dot{\phi}}\dot{\varphi}
\right)^2- \frac{1}{24}\partial_i \beta \, \partial^i \beta \nonumber \\
& & +\frac{1}{6} \nabla^2
 \left\{ \frac{1}{8}\beta^2+
\frac{3}{2}\frac{\partial^i \partial^j}{(\nabla^2)^2}\left[-\frac{1}{4}\beta
 D_{ij} \beta +2 \psi D_{ij} E + \frac{1}{H}\psi D_{ij} \dot{E}\right]
\right\}\,.
\label{R_second_gen}
\end{eqnarray}
This quantity is the true GI generalization of ${\cal R}^{(2)}$ computed 
in \cite{ABMR}, which also takes into account infinitesimal spatial 
(scalar) transformation in addition to the infinitesimal time translations 
considered in \cite{ABMR}.

\subsection{The comoving intrinsic curvature potential perturbation $R_C$}

In order to introduce the third curvature (potential) perturbation, 
let us first consider 
the intrinsic 3-curvature $ ^{(3)}R$ associated with a foliation of space-time 
with constant proper time $t$ for a flat Universe.
Using eq.(49) and eq.(58) of \cite{NohHwang} we obtain:
\begin{eqnarray}
^{(3)}R &=& \frac{1}{a^2}\left[2 \partial^i\partial_j C_i^j-2 \partial^i
\partial_i C_j^j +4 C^{i\,j}\left(-2 \partial_j\partial_k C_i^k+
\partial_k\partial^k C_{i\,j}+\partial_i\partial_j C_k^k\right)
-\left(2 \partial_k C_j^k -\partial_j C_k^k\right)
\left(2 \partial_i C^{i\,j} -\partial^j C_i^i\right) \right. \nonumber \\
& & \left.
+\partial^k C^{i\,j}\left(3 \partial_k C_{i\,j}-2 \partial_j C_{i\,k}
\right) \right]
\label{def3RconC}
\end{eqnarray}
with 

$$
C_{i\,j}=-\delta_{i\,j}
\left(\psi+\psi^{(2)}\right)+\frac{1}{2} D_{i\,j}
\left(E+E^{(2)}\right) \,.
$$
In our notations the intrinsic curvature is:
\begin{eqnarray}
^{(3)}R &=& ^{(3)}R^{(1)} + ^{(3)}R^{(2)} \nonumber \\
&=& \frac{4}{a^2} \nabla^2 \left( \psi+\frac{1}{6}
 \nabla^2 E\right)+ \frac{4}{a^2}\left\{ \nabla^2 \left( \psi^{(2)}
+\frac{1}{6} \nabla^2
E^{(2)}\right) +4 \left( \psi+\frac{1}{6} \nabla^2 E\right)  
\nabla^2 \left( \psi+\frac{1}{6} \nabla^2 E\right) \right.
\nonumber \\
& & \left. +\frac{3}{2}
\partial^i \left( \psi+\frac{1}{6} \nabla^2 E\right)\partial_i
\left( \psi+\frac{1}{6} \nabla^2 E\right)
-\frac{1}{2}  
\nabla^2 E \nabla^2 \left( \psi+\frac{1}{6} \nabla^2 E\right)
\right. \nonumber \\
& & \left.
-\frac{1}{2}\partial^i\partial^j E \partial_i\partial_j
\left( \psi+\frac{1}{6} \nabla^2 E\right)-\frac{1}{2} \partial^i
\left( \psi+\frac{1}{6} \nabla^2 E\right)\partial_i 
\nabla^2 E
-\frac{1}{16} \partial^i\nabla^2 E \partial_i 
\nabla^2 E
\right. \nonumber \\
& & \left.
+\frac{1}{16}\partial^i\partial^j\partial^k E 
\partial_i\partial_j\partial_k E\right\}\,.
\label{3curvature}
\end{eqnarray}
Instead of using Eq. (\ref{CP_order2_linear}), we now consider the potential 
$\hat{\psi}^{(2)}_C$ of the intrinsic curvature to second order
$^{(3)}R^{(2)}$, given by 
$^{(3)}R^{(2)}=\frac{4}{a^2} \nabla^2 \hat{\psi}^{(2)}_C$. 
We then obtain the comoving GI expression by going from a general slicing to a
comoving slicing:
\begin{eqnarray}
R^{(2)}_C&=&\psi^{(2)}
+\frac{H}{\dot{\phi}}\varphi^{(2)}+\frac{1}{6}\nabla^2 E^{(2)}-
\frac{\varphi}{\dot{\phi}}\left( \dot{\psi}+2 H \psi \right)-
\frac{H}{\dot{\phi}^2}\varphi \dot{\varphi}+\frac{\varphi^2}{\dot{\phi}^2}
\left( -H^2 -\frac{\dot{H}}{2}+\frac{H}{2}\frac{\ddot{\phi}}{\dot{\phi}}
\right) \nonumber \\
& & + \frac{1}{6 a^2}\frac{1}{\dot{\phi}^2}\left[\partial^i\varphi
\partial_i\varphi-\frac{1}{2}\nabla^2\left(\varphi^2\right)\right]-
\frac{1}{6 a}\frac{1}{\dot{\phi}}\left[\partial^i\beta
\partial_i\varphi-\frac{1}{2}\nabla^2\left(\beta \varphi \right)\right]+
\frac{1}{4}
\frac{\partial^i \partial^j}{\nabla^2}\left[-2 H 
\frac{\varphi}{\dot{\phi}}
 D_{ij} E  \right. \nonumber \\
& & \left.
- \frac{\varphi}{\dot{\phi}} D_{ij} \dot{E}
+ \frac{1}{a^2}  \frac{\varphi}{\dot{\phi}^2}
D_{ij} \varphi -
 \frac{1}{2a} \frac{1}{\dot{\phi}} \left(
\varphi D_{ij} \beta + \beta  D_{ij} \varphi
\right) \right]
+\frac{1}{\nabla^2}
\left[
4 \left( \psi+\frac{1}{6} \nabla^2 E+\frac{H}{\dot{\phi}}\varphi 
\right)  \right.
\nonumber \\
& & \left.
\nabla^2 \left( \psi
+\frac{1}{6} \nabla^2 E
+\frac{H}{\dot{\phi}}
\varphi \right)+ 
\frac{3}{2}
\partial^i \left( \psi+\frac{1}{6} \nabla^2 E
+\frac{H}{\dot{\phi}}\varphi\right)\partial_i
\left( \psi+\frac{1}{6} \nabla^2 E+\frac{H}{\dot{\phi}}\varphi
\right) 
\right.
\nonumber \\
& & \left. 
-\frac{1}{2}  
\nabla^2 E \nabla^2 \left( \psi+\frac{1}{6} \nabla^2 E
+\frac{H}{\dot{\phi}}\varphi\right)
-\frac{1}{2}\partial^i\partial^j E \partial_i\partial_j
\left( \psi+\frac{1}{6} \nabla^2 E+\frac{H}{\dot{\phi}}\varphi
\right)
\right. \nonumber \\
& & \left.
-\frac{1}{2} \partial^i
\left( \psi+\frac{1}{6} \nabla^2 E+\frac{H}{\dot{\phi}}\varphi
\right)\partial_i 
\nabla^2 E
-\frac{1}{16} \partial^i\nabla^2 E \partial_i 
\nabla^2 E
\right. \nonumber \\
& & \left.
+\frac{1}{16}\partial^i\partial^j\partial^k E 
\partial_i\partial_j\partial_k E\right]
\label{ComovingRord2full}
\end{eqnarray}

\subsection{Comparison among curvature perturbations in the long-wavelength 
limit}

The preceding three expressions undergo a considerable simplification in the 
long-wavelength limit.
For large scales we obtain:
\be
R^{(2)}_A=\psi^{(2)}
+\frac{H}{\dot{\phi}}\varphi^{(2)}-
\frac{\varphi}{\dot{\phi}}\left( \dot{\psi}+2 H \psi \right)-
\frac{H}{\dot{\phi}^2}\varphi \dot{\varphi}+\frac{\varphi^2}{\dot{\phi}^2}
\left( -H^2 -\frac{\dot{H}}{2}+\frac{H}{2}\frac{\ddot{\phi}}{\dot{\phi}}
\right)
\label{ComovingRord2_LW_limit}
\ee

\be
R^{(2)}_B = 
 \psi^{(2)} + \frac{H}{\dot{\phi}} \varphi^{(2)} +
\frac{1}{2}\left ( 2 H^2 +\dot{H}-H\frac{\ddot{\phi}}{\dot{\phi}}
\right)^{-1}\left( \dot{\psi}+2 H \psi +\frac{H}{\dot{\phi}}\dot{\varphi}
\right)^2
\label{R_second_gen_LW_limit}
\ee

\begin{eqnarray}
R^{(2)}_C&=&\psi^{(2)}
+\frac{H}{\dot{\phi}}\varphi^{(2)}-
\frac{\varphi}{\dot{\phi}}\left( \dot{\psi}+2 H \psi \right)-
\frac{H}{\dot{\phi}^2}\varphi \dot{\varphi}+\frac{\varphi^2}{\dot{\phi}^2}
\left( -H^2 -\frac{\dot{H}}{2}+\frac{H}{2}\frac{\ddot{\phi}}{\dot{\phi}}
\right) \nonumber \\
& &
+\frac{1}{\nabla^2}\left[ 4 \left( \psi+\frac{H}{\dot{\phi}}\varphi 
\right)  \nabla^2 \left( \psi +\frac{H}{\dot{\phi}}
\varphi \right)+ 
\frac{3}{2}
\partial^i \left( \psi
+\frac{H}{\dot{\phi}}\varphi\right)\partial_i
\left( \psi+\frac{H}{\dot{\phi}}\varphi
\right)\right]
\label{ComovingRord2full_LW_limit}
\end{eqnarray}
in this limit the difference between $R^{(2)}_A$ and $R^{(2)}_B$ is given by
\cite{vernizzi}:
\be
R^{(2)}_B-R^{(2)}_A=\frac{1}{2} 
\frac{\left(\dot{R}^{(1)}+2 H R^{(1)}\right)^2}{\left ( 2 H^2 +\dot{H}-H\frac{\ddot{\phi}}{\dot{\phi}}
\right)}\,.
\label{differenceR}
\ee
Let us note that, as for the first order $R^{(1)}$, $R^{(2)}_A$ also is
constant in time in this limit \cite{LMS,vernizzi}.

Further on using different parametrizations of the metric many other gauge 
invariant quantities can be constructed. For example from \cite{LythRodriguez}
one can define another variable $R^{(2)}_{LR}$, which is related to 
$R^{(2)}_A$ on large scales by
\be
R^{(2)}_A=R^{(2)}_{LR}- \left(R^{(1)}\right)^2 \,.
\label{R_LR}
\ee
We also consider $R^{(2)}_C$ since it is directly related to a 
geometrical quantity: the intrinsic 3-curvature.

\section{Long-wavelength Second Order Curvature Perturbations in UCG}

As in one of our previous papers on 
back-reaction \cite{FMVV_2}, we choose to work in the uniform curvature 
gauge (UCG), which can be generalized straightforwardly to second order
through the conditions: $\psi=\psi^{(2)}=0$ and $E=E^{(2)}=0$. 
In this 
gauge, the evolution equation for inflaton fluctuations is regular also 
during 
the coherent oscillation of the scalar field \cite{FB}. 
In the UCG and in the long wavelenght limit we have, for the given
second order curvature perturbation potentials, the following results:
 
\be
R^{(2)}_A=\frac{H}{\dot{\phi}}\varphi^{(2)}-
\frac{H}{\dot{\phi}^2}\varphi \dot{\varphi}+\frac{H^2}{\dot{\phi}^2}
\varphi^2
\left( -1 -\frac{\dot{H}}{2 H^2}+\frac{1}{2 H}\frac{\ddot{\phi}}{\dot{\phi}}
\right)
\label{ComovingRord2_LW_limit_UCG}
\ee

\be
R^{(2)}_B = 
 \frac{H}{\dot{\phi}} \varphi^{(2)} +
\frac{1}{2}\left ( 2 H^2 +\dot{H}-H\frac{\ddot{\phi}}{\dot{\phi}}
\right)^{-1} \frac{H^2}{\dot{\phi}^2}\dot{\varphi}^2
\label{R_second_gen_LW_limit_UCG}
\ee

\be
R^{(2)}_C = \frac{H}{\dot{\phi}}\varphi^{(2)}-
\frac{H}{\dot{\phi}^2}\varphi \dot{\varphi}+\frac{H^2}{\dot{\phi}^2}
\varphi^2
\left( -1 -\frac{\dot{H}}{2 H^2}+\frac{1}{2 H}\frac{\ddot{\phi}}{\dot{\phi}}
\right)
+\frac{1}{\nabla^2}\left[ 4 \frac{H^2}{\dot{\phi}^2}\varphi 
\nabla^2
\varphi+ 
\frac{3}{2} \frac{H^2}{\dot{\phi}^2}
\partial^i
\varphi \partial_i
\varphi
\right] \,.
\label{ComovingRord2full_LW_limit_UCG}
\ee

On subtracting the average of the quadratic piece (in order to also have  
a zero average for the non-linear piece), it is useful to introduce the 
parameter of nonlinearity $f_{NL}$ \cite{LythRodriguez}:
\be 
R=R_{\rm L}+\frac{3}{5} f_{\rm NL}\left[R_{\rm L}^2-\langle R_{\rm L}^2 
\rangle \right]
\label{formNLR}
\ee
where $R_{\rm L}$ is the Gaussian part of $R$ (that is the first order part
$R^{(1)}$).
Using the first order comoving curvature perturbation together with the 
first two second order definitions given in Eqs. 
(\ref{ComovingRord2_LW_limit_UCG}), (\ref{R_second_gen_LW_limit_UCG}), 
we obtain for a $\frac{m^2}{2} \phi^2$ chaotic inflation and 
using the approximation $\dot{\varphi}=-\frac{\dot{H}}{H} \varphi$ and 
Eq.(\ref{relation_ord1_ord2}),
to leading order in the slow-roll parameter 
$\epsilon=-\frac{\dot{H}}{H^2}$, the following results for $f_{NL}$:
$$
f^A_{\rm NL}=-\frac{5}{3}
$$
$$
f^B_{\rm NL}=\frac{5}{6}\epsilon \,.
$$
Let us note that we cannot evaluate $f^C_{\rm NL}$ in analogy with 
\cite{LythRodriguez} because the non-local terms are not subleading in the 
long-wavelength approximation. 

In the next two sections we proceed with the evaluation of the spectrum 
of the non-linear corrections to the curvature $R_A$ and of the second order 
GI field fluctuation $Q^{(2)}$ defined in \cite{Malik}. 
The evaluation of the quantum correlator of second-order GI variables 
(see also \cite{weinberg}) 
involves a sum over momenta, which is 
plagued by ultraviolet divergencies. 
The adiabatic subtraction used in the 
calculation of the finite part of energy and pressure of 
fluctuations \cite{FMVV_I,FMVV_2} is roughly equivalent to only considering 
super-Hubble fluctuations. Motivated by this analogy we shall also use this 
criterium in the present case: all the integrals which follow have 
the physical Hubble radius as an upper limit of integration. 
The same integrals also need a lower limit of integration 
which defines the region of nearly scale invariant (red tilted) spectra
generated by inflation: 
we choose this 
lower limit of integration, a comoving momentum $\ell$, of order of the
Hubble rate when inflation begins, $H_0$, as previously  found 
in analytical \cite{VF} and numerical investigation \cite{FMVV_I} 
(see also  \cite{AF}). 
At the present state of our knowledge (i.e. without
any theoretical input for the choice of $\ell$), we believe that this is a
natural choice as compared to considering $\ell$ of the order of the present
comoving Hubble radius as is done in \cite{Lyth}.
From now on we shall restrict ourselves to the slow-roll stage of 
$\frac{m^2}{2} \phi^2$ chaotic inflation, during which 
$\dot{H}\simeq -m^2/3$ holds.

\subsection{Analytic Evaluation of curvature $R^{(2)}_A$}

Within the above approximation from eq.(\ref{ComovingRord2_LW_limit_UCG}), 
to leading order in the slow-roll parameter
and using Eq.(28) of our \cite{FMVV_2}, one obtains
\be
\frac{H}{\dot{\phi}}\varphi^{(2)}\sim \epsilon R^{(1) \,2}
\label{andamento_varphi2}
\ee
and
\be
R^{(2)}_A=R^{(1)\, 2}\left[-1+{\cal O}\left(\epsilon\right)\right]\,.
\label{R2A_leading}
\ee
From the above equation and Eq. (\ref{R_LR}) it is evident 
that $R_{LR}$ is of order a slow-roll parameter times the square of the 
first-order perturbation.

Let us consider the quantum correlation function for this second order gauge 
invariant variable
\begin{eqnarray}
\langle 0 |R^{(2)}_A({\bf x}) R^{(2)}_A({\bf y})| 0 \rangle&=& 
\left(1+{\cal O}\left(\epsilon\right)\right) \langle 0 
|\left(R^{(1)}({\bf x})\right)^2 \left(R^{(1)}({\bf y})\right)^2
| 0 \rangle \nonumber \\
&\simeq&\frac{H^4}{\dot{\phi}^4} 
 \langle 0 |\left(\varphi(t, {\bf x})\right)^2 \left(\varphi(t, {\bf y})
\right)^2| 0 \rangle \,.
\label{CF_R2A}
\end{eqnarray}

The quantized scalar field variable $\varphi$ is given by
\be
\hat{\varphi}  (t, {\bf x}) = \frac{1}{(2 \pi)^{3/2}} \int d{\bf k}
\left[ \varphi_{k} (t) \, e^{i {\bf k} \cdot {\bf x}} \,\, \hat{a}_{\bf k} 
+ \varphi_{k}^* (t) e^{- i {\bf k} \cdot {\bf x}} \,\,
\hat{a}^\dagger_{{\bf k}} \right]
\label{quantumFourier_varphi}
\ee
in the slow-roll approximation and on large scales (but $< 2 \pi/\ell$) 
we consider 
\cite{FMVV_2}
\be
\varphi_k (t)=-\frac{i}{H} \frac{H(t_k)^2}{\sqrt{2 k^3}} \,,
\label{varphi_infrared}
\ee
with 
\be
H(t_k) \simeq H_0 \left(1-2 \epsilon_0 \ln \frac{k}{H_0}\right)^{1/2}
\label{Hcrossing}
\ee
which is the Hubble parameter when the fluctuations crosses the horizon 
and $\epsilon_0=\epsilon(t=0)$. 
The power spectrum of first order curvature perturbation, at large scales
and in the UCG,  is therefore:
\be
P_{R^{(1)}} (k) \equiv \frac{k^3}{2 \pi^2} |R^{(1)}_k|^2 = 
\frac{k^3}{2 \pi^2} \frac{H^2}{{\dot \phi}^2}
|\varphi_{k} (t)|^2 = \frac{3}{8 \pi^2} 
\frac{H(t_k)^4}{m^2 M_{pl}^2} \,.
\label{1storderspectrum}
\ee 
It is important to note that by using the expression (\ref{Hcrossing}) for 
$H(t_k)$ we obtain a power spectrum of curvature perturbations in 
(\ref{1storderspectrum}) having the correct value to leading order of the 
spectral index and exhibiting running during slow-roll.

The physical content of the correlation function for $R^{(2)}_A$ is 
conveniently described by the power spectrum 
associated with its fourier transform
\be
\langle 0 |R^{(2)}_A({\bf x}) R^{(2)}_A({\bf y})| 0 \rangle
=\frac{1}{(2 \pi)^3} \int d^3 k \e^{i {\bf k} \cdot ({\bf x}-{\bf y})}
|R^{(2)}_{A\,k}(t)|^2 \,
\label{Corr_RA}
\ee
which becomes, using Eq.(\ref{CF_R2A})
\begin{eqnarray}
& & \langle 0 |R^{(2)}_A({\bf x}) R^{(2)}_A({\bf y})| 0 \rangle \simeq 
\frac{H^4}{4 \dot{H}^2}\frac{1}{M_{pl}^4}
\frac{1}{(2 \pi)^6}
\left\{  \int
d^3 k_1 d^3 k_2 \frac{H(t_{k_1})^4}{k_1^3}\frac{H(t_{k_2})^4}{k_2^3}
\frac{1}{2 H^4}\Theta(a H-k_1)
 \right. \nonumber \\ 
& &\,\,\,\,\, \,\,\,\,\,\left.
\Theta(a H-k_2)
\Theta(k_1-l)
\Theta(k_2-l)
\exp^{i ({\bf k_1}+{\bf k_2}) \cdot ({\bf x}-{\bf y})}+\int d^3 k_1 d^3 k_2
 \frac{H(t_{k_1})^4}{k_1^3}\frac{H(t_{k_2})^4}{k_2^3}\frac{1}{4 H^4}
\right. \nonumber\\
& &\,\,\,\,\,\,\,\,\,\, \left. \Theta(a H-k_1) \Theta(a H-k_2)
\Theta(k_1-l)
\Theta(k_2-l)
   \right\} \,,
\label{Corr_R2A_varphi}
\end{eqnarray}
where we have considered only modes which are super-Hubble, but greater then 
comoving infrared cut-off $l$, using the Heaviside functions $\Theta$.
We can perform a Fourier transform with respect to ${\bf r}={\bf x}-{\bf y}$
of  Eq.(\ref{Corr_R2A_varphi}) and obtain

\begin{eqnarray}
|R^{(2)}_{A\,p}(t)|^2&\simeq&
\frac{H^4}{4 \dot{H}^2}\frac{1}{M_{pl}^4}
\frac{1}{(2 \pi)^3}\int d^3 k \frac{H(t_{k})^4}{k^3}\frac{H(t_{p-k})^4}{
|{\bf p}-{\bf k}|^3}\frac{1}{2 H^4}
\Theta(a H-k)\Theta(a H-|{\bf p}-{\bf k}|) \nonumber \\ 
& & \Theta(k-l)
\Theta(|{\bf p}-{\bf k}|-l)+ (2 \pi)^3 \delta^{(3)}(p) \tilde{Q}_0 
\label{trasf_fourier_R2A}
\end{eqnarray}
where
\be
\tilde{Q}_0=
\frac{H^4}{4 \dot{H}^2}\frac{1}{M_{pl}^4}
\frac{1}{(2 \pi)^6}
\int d^3 k_1 d^3 k_2
 \frac{H(t_{k_1})^4}{k_1^3}\frac{H(t_{k_2})^4}{k_2^3}\frac{1}{4 H^4}
 \Theta(a H-k_1) \Theta(a H-k_2)
\Theta(k_1-l)
\Theta(k_2-l)\,.
\label{Q_0_R2A}
\ee
For $p \ne 0$ and on evaluating the integral in 
Eq.(\ref{trasf_fourier_R2A}) to leading order and in the momentum window given 
by the condition $l<<p<<a(t) H(t)$ one obtains the
expression:
\be
|R^{(2)}_{A\,p}(t)|^2\simeq
\frac{1}{32 \pi^2} \frac{1}{M_{pl}^4}
\frac{H_0^4}{\epsilon_0^2}
\frac{1}{p^3}
\left\{ \sum_{i=0}^3 A_i \left(\ln{\frac{p}{l}}\right)^i
+g\left(\frac{l}{p}\right)+h\left(\frac{p}{a(t)H(t)}\right)
\right\} 
\label{R2A_value}
\ee 
which is time independent in our approximation.
The coefficients $A_i$ are given in appendix B,
$g\left(\frac{l}{p}\right)={\cal O} \left(\left(\ln{\frac{p}{l}}\right)^2 
\frac{l}{p}\right)$ and $h\left(\frac{p}{a(t)H(t)}\right)=
{\cal O} \left(\left(\ln{\frac{a(t)H(t)}{p}}\right)^4
\left(\frac{p}{a(t)H(t)}\right)^3\right)$.

Substituting for the coefficients $A_i$ we obtain the following result:
\begin{eqnarray}
|R^{(2)}_{A\,p}|^2 &\simeq&
\frac{1}{32 \pi^2} \frac{1}{M_{pl}^4}
\frac{1}{p^3}
\left\{-2.8405\frac{1}{\dot{H}} H(t_p)^6+19.3975  H(t_p)^4-
14.9341 \dot{H} H(t_p)^2+ 9.7273 \dot{H}^2+4 \frac{1}{\dot{H}^2}
H(t_p)^8 \right. \nonumber \\
& & \left. \ln{\frac{p}{l}}+16  H(t_p)^4 \left(\ln{\frac{p}{l}}\right)^2
+\frac{16}{3}  H(t_p)^4 \left(\ln{\frac{p}{l}}\right)^3
+g\left(\frac{l}{p}\right)+h\left(\frac{p}{a(t)H(t)}\right)
\right\} 
\,.
\label{R2A_value_developed}
\end{eqnarray}
On taking only the leading term for the coefficents of the different 
powers in $\ln(p/l)$, we obtain:
\begin{eqnarray}
|R^{(2)}_{A\,p}|^2 &\simeq& \frac{1}{32 \pi^2} \frac{1}{M_{pl}^4} 
\frac{H(t_p)^4}{p^3} \left\{ 
8.5215 \frac{H(t_p)^2}{m^2} + 36 \frac{H(t_p)^4}{m^4} 
\ln{\frac{p}{l}} + 16  \left(\ln{\frac{p}{l}}\right)^2
+ \frac{16}{3}  \left(\ln{\frac{p}{l}}\right)^3 \right\} \nonumber \\
&\simeq&
\frac{1}{24 \pi^2} |R^{(1)}_p|^2 
\left\{ 
8.5215 \, \frac{H(t_p)^2}{M_{pl}^2} + 36 \frac{H(t_p)^4}{m^2 M_{pl}^2} 
\left( \ln{\frac{p}{l}} \right) + 16 \, 
\frac{m^2}{M_{pl}^2} \left(\ln{\frac{p}{l}}\right)^2
+ \frac{16}{3} \, \frac{m^2}{M_{pl}^2} 
\left(\ln{\frac{p}{l}}\right)^3 \right\}
\,,
\label{R2A_value_developed_lead}
\end{eqnarray}
where in the last equality we have used Eq.(\ref{1storderspectrum}). 
This last expression shows compactly one of the main results of our 
investigation: the spectrum of this second order GI curvature perturbation 
is proportional to the first order spectrum through logarithmic corrections 
which encode the scale $l$ at which inflation started. 
Without these logarithmic corrections the second order curvature 
perturbation would be completely local 
and much smaller than first order 
terms since for the scales of interest $H(t_k) << M_{pl}$.


\subsection{Analytic Evaluation of $Q^{(2)}$}

Let us consider the second order GI scalar field fluctuation $Q^{(2)}$ 
(which is a possible second order generalization of the first order Mukhanov
variable \cite{mukhanov})  defined in \cite{Malik} as the second order field
fluctuations on uniform curvature hypersurfaces in the longwavelength limit.
The uniform curvature hypersurfaces is defined as the slicing with
$\hat{\psi}=\hat{\psi}^{(2)}=0$, and one obtains the following result 
\cite{MW,Malik}:

\be 
Q^{(2)}=\varphi^{(2)}+\frac{\dot{\phi}}{H} \psi^{(2)}+\frac{1}{H}\psi 
\dot{\varphi}+\left(1+\frac{1}{2H}\frac{\ddot{\phi}}{\dot{\phi}}-\frac{1}{2}
\frac{\dot{H}}{H^2}\right) \frac{\dot{\phi}}{H} \psi^2+\frac{\dot{\phi}}{H^2}
\psi \dot{\psi}\,.
\label{Q_Malik}
\ee

In the UCG, as for the first order, this GI variable is simply the second
order field fluctuations $\varphi^{(2)}$, studied in our previous paper 
\cite{FMVV_2}.
Thus the equation of motion for this second order GI field fluctuation in the 
UCG, to leading order in the slow-roll parameter and in the 
long-wavelength approximation is given by \cite{FMVV_2}:

\be
\ddot{Q}^{(2)} + 3 H \dot{Q}^{(2)}+ 3 \dot H Q^{(2)} = m^2
\frac{\dot{\phi}}{2 H} \frac{\varphi^2}{M_{pl}^2} \,.
\label{Eq_Q_scalata_second_leading_order_approximate}
\ee

It is interesting to comment on the self-interaction mediated by gravity 
corrected for the inflaton. Just as the feedback of metric perturbation 
vanishes in first order perturbation theory for $\dot{\phi}=0$ 
(see Eq. (7) of \cite{FMVV_2}), it also vanishes for the 
self-interaction to second order 
in Eq. (\ref{Eq_Q_scalata_second_leading_order_approximate}).  

Integrating the above in two steps we obtain:
\be
\dot{Q}^{(2)}+ 3 H Q^{(2)} = m^2 
\int  dt' \frac{\dot{\phi}}{2 H(t')} \frac{\varphi(t', {\bf x})^2}{M_{pl}^2} 
\,,
\label{Eq_Q_scalata_second_leading_order_approximate_int}
\ee
and
\be
Q^{(2)} (t, {\bf x}) =\frac{1}{a(t)^3} \int^t dt' a(t')^3
\int^{t'} dt'' \frac{\dot{\phi}}{2 H(t'')}m^2 
\frac{\varphi(t'', {\bf x})^2}{M_{pl}^2} \,.
\label{Eq_Q_scalata_Green_Function}
\ee

Let us consider the quantum
correlation function of this second order $Q^{(2)}$ gauge invariant variable
 $\langle 0 |Q^{(2)}(t, {\bf x}) Q^{(2)}(t, {\bf y})| 0 \rangle$.
The quantized scalar variable $Q^{(2)}$ is 
given 
by
\begin{eqnarray}
\hat{Q}^{(2)}  (t, {\bf x}) &=& \frac{1}{(2 \pi)^{3/2}} \int d{\bf k}
\left[ Q^{(2)}_{k} (t) \, e^{i {\bf k} \cdot {\bf x}} \,\, \hat{b}_{\bf k} 
+ Q^{(2)*}_{k} (t) e^{- i {\bf k} \cdot {\bf x}} \,\,
\hat{b}^\dagger_{{\bf k}} \right]\,.
\label{quantumFourier_q}
\end{eqnarray}

The physical content of the correlation function for $Q^{(2)}$ is
better described by the power spectrum 
associated with its fourier transform
\be
\langle 0 |Q^{(2)}(t, {\bf x}) Q^{(2)}(t, {\bf y})| 0 \rangle
=\frac{1}{(2 \pi)^3} \int d^3 k \e^{i {\bf k} \cdot ({\bf x}-{\bf y})}
|Q^{(2)}_k (t)|^2
\label{Corr_q}
\ee
and using 
Eq.(\ref{Eq_Q_scalata_Green_Function}) and Eq.(\ref{varphi_infrared}), 
one obtains

\begin{eqnarray}
& & \langle 0 |Q^{(2)}(t, {\bf x}) Q^{(2)}(t, {\bf y})| 0 \rangle = 
\frac{1}{a(t)^6} \left(\frac{\dot{\phi}}{2}\frac{m^2}{M_{pl}^2}\right)^2
\frac{1}{(2 \pi)^6}
\left\{  \int
d^3 k_1 d^3 k_2 \frac{H(t_{k_1})^4}{k_1^3}\frac{H(t_{k_2})^4}{k_2^3}
\frac{1}{2} \left[ \int^t dt' a(t')^3
\right. \right. \nonumber \\
& &\,\,\,\,\, \,\,\,\,\,\left. \left.
\int^{t'} dt'' 
\frac{1}{H(t'')^3}
\Theta(a(t'') H(t'')-k_1) \Theta(a(t'') H(t'')-k_2)\right]^2
\Theta(k_1-l)
\Theta(k_2-l)
\exp^{i ({\bf k_1}+{\bf k_2}) \cdot ({\bf x}-{\bf y})}
\right. \nonumber \\ 
& &\,\,\,\,\, \,\,\,\,\,\left.
+\int d^3 k_1 d^3 k_2
 \frac{H(t_{k_1})^4}{k_1^3}\frac{H(t_{k_2})^4}{k_2^3}\frac{1}{4}
\left[ \int^t dt' a(t')^3 \int^{t'} dt'' 
\frac{1}{H(t'')^3}
\Theta(a(t'') H(t'')-k_1) \right]
 \right. \nonumber \\ 
& &\,\,\,\,\, \,\,\,\,\,\left.
\left[ \int^t dt' a(t')^3 \int^{t'} dt'' 
\frac{1}{H(t'')^3}
\Theta(a(t'') H(t'')-k_2) \right]
\Theta(k_1-l)
\Theta(k_2-l)
\right\}\,.
\label{Corr_q_varphi_Green}
\end{eqnarray}

On performing the Fourier transformation w.r.t. ${\bf r}={\bf x}-{\bf y}$
of  Eq.(\ref{Corr_q_varphi_Green}) 
one finds
\begin{eqnarray}
|Q^{(2)}_p(t)|^2&=&\frac{1}{a(t)^6} \left(\frac{\dot{\phi}}{2}
\frac{m^2}{M_{pl}^2}\right)^2
\frac{1}{(2 \pi)^3}\int d^3 k \frac{H(t_{k})^4}{k^3}\frac{H(t_{p-k})^4}{
|{\bf p}-{\bf k}|^3}\frac{1}{2}
\left[ \int^t dt' a(t')^3 \int^{t'} dt'' 
\frac{1}{H(t'')^3}
\Theta(a(t'') H(t'')-k) \right.
\nonumber \\ 
& & \left.
\Theta(a(t'') H(t'')-|{\bf p}-{\bf k}|)
\right]^2
 \Theta(k-l)
\Theta(|{\bf p}-{\bf k}|-l)+ (2 \pi)^3 \delta^{(3)}(p) Q_0 
\label{trasf_fourier_Green}
\end{eqnarray}
where
\begin{eqnarray}
Q_0 &=& \frac{1}{a(t)^6}
\left(\frac{\dot{\phi}}{2}\frac{m^2}{M_{pl}^2}\right)^2 \frac{1}{(2 \pi)^6}
\int d^3 k_1 d^3 k_2
 \frac{H(t_{k_1})^4}{k_1^3}\frac{H(t_{k_2})^4}{k_2^3}\frac{1}{4}
\left[ \int^t dt' a(t')^3 \int^{t'} dt'' 
\frac{1}{H(t'')^3}
\Theta(a(t'') H(t'')-k_1) \right]
\nonumber \\ 
& &
\left[ \int^t ds' a(s')^3 \int^{s'} ds'' 
\frac{1}{H(s'')^3}
\Theta(a(s'') H(s'')-k_2) \right]
\Theta(k_1-l)
\Theta(k_2-l) \,.
\label{Q_0_Green}
\end{eqnarray}
On considering $p \ne 0$ and replacing the $\Theta$ function by
the domain of integration, it is possible to perform the time integrals 
exactly:
\begin{eqnarray}
|Q^{(2)}_p(t)|^2&=& \left(\frac{\dot{\phi}}{2}
\frac{m^2}{M_{pl}^2}\right)^2
\frac{1}{(2 \pi)^2}\frac{1}{2} 
\int_{l}^{a(t)H(t)} d k
\int_{-1}^{+1} d y \Theta(a(t) H(t)-(p^2+k^2-2 k p y)^{1/2} )
\Theta((p^2+k^2-2 k p y)^{1/2} -l)
\nonumber \\
& &
\frac{1}{k}\frac{1}{(p^2+k^2-2 k p y)^{3/2}}H_0^6 \left( 1+2 
\frac{\dot{H}}{H_0^2}\ln \frac{k}{H_0}\right)^2 
\left( 1+2 
\frac{\dot{H}}{H_0^2}\ln \frac{(p^2+k^2-2 k p y)^{1/2}}{H_0}\right)^2 
\nonumber \\
& &
\frac{1}{36 \dot{H}^2}
\e^{-\frac{3}{2}\frac{H(t)^2}{\dot{H}}} \left\{\frac{1}{f(k, p, y)^2}
\left(-\frac{3}{2 \dot{H}}\right)^{1/2}
\left[ \Gamma\left(\frac{1}{2}, -\frac{3}{2}\frac{H(t)^2}{\dot{H}}\right)-
\Gamma\left(\frac{1}{2}, -\frac{3}{2}
\frac{f(k, p, y)^2}{\dot{H}}\right)\right]
\right.
\nonumber \\ 
& & \left. -\left(-\frac{3}{2 \dot{H}}\right)^{3/2}
\left[ \Gamma\left(-\frac{1}{2}, -\frac{3}{2}\frac{H(t)^2}{\dot{H}}\right)-
\Gamma\left(-\frac{1}{2}, -\frac{3}{2}
\frac{f(k, p, y)^2}{\dot{H}}\right)\right]
\right\}^2 \,,
\label{trasf_fourier_Green_numerical}
\end{eqnarray}
where $f(k, p, y)$ is given by 
$$
f(k, p, y)=Min \left[ H_0 \left( 1+2 
\frac{\dot{H}}{H_0^2}\ln \frac{k}{H_0}\right)^{1/2}, H_0 \left( 1+2 
\frac{\dot{H}}{H_0^2}\ln \frac{(p^2+k^2-2 k p y)^{1/2}}{H_0}\right)^{1/2}
\right] \,.
$$

It is also possible to obtain an explicit approximate espression for the
integral starting from Eq.(\ref{trasf_fourier_Green}) subject to the condition
$l<<p<<a(t) H(t)$ and the approximation  $l<<p<<a(t'') H(t'')$.
In particular we find
\begin{eqnarray}
& & |Q^{(2)}_p(t)|^2 \simeq \frac{1}{a(t)^6} \left(\frac{\dot{\phi}}{2}
\frac{m^2}{M_{pl}^2}\right)^2
\frac{1}{8 \pi^2} H_0^2 \frac{1}{p^3}
\left\{ \sum_{i=0}^3 A_i \left(\ln{\frac{p}{l}}\right)^i
+g\left(\frac{l}{p}\right)+h\left(\frac{p}{a(t)H(t)}\right)
\right\} 
\frac{1}{36}\frac{1}{\dot{H}^2}
e^{-3\frac{H_0^2}{\dot{H}}} 
\nonumber \\ 
& &\!\!\!\!\!\!
\left\{\left(-\frac{3}{2}\frac{H_0^2}{\dot{H}}\right)^{1/2} 
\left[\Gamma\left(\frac{1}{2},-\frac{3}{2}\frac{H^2}{\dot{H}}\right)
-\Gamma\left(\frac{1}{2},-\frac{3}{2}\frac{H_0^2}{\dot{H}}\right)\right]
-\left(-\frac{3}{2}\frac{H_0^2}{\dot{H}}\right)^{3/2} 
\left[\Gamma\left(-\frac{1}{2},-\frac{3}{2}\frac{H^2}{\dot{H}}\right)
-\Gamma\left(-\frac{1}{2},-\frac{3}{2}\frac{H_0^2}{\dot{H}}\right)\right]
\right\}^2
\label{q2_value_Green}
\end{eqnarray}
where, as in the previous subsection, 
the coefficients $A_i$ are discussed in the appendix,
$g\left(\frac{l}{p}\right)={\cal O} \left(\left(\ln{\frac{p}{l}}\right)^2 
\frac{l}{p}\right)$ and $h\left(\frac{p}{a(t)H(t)}\right)=
{\cal O} \left(\left(\ln{\frac{a(t)H(t)}{p}}\right)^4
\left(\frac{p}{a(t)H(t)}\right)^3\right)$.


Let us also investigate a much more crude approximation, obtained 
on neglecting $\dot{Q}^{(2)}$ in 
Eq.(\ref{Eq_Q_scalata_second_leading_order_approximate_int}) and 
considering the infrared limit $\dot{\varphi}=-\frac{\dot{H}}{H}\varphi$,
which leads to the expression 

\be
Q^{(2)} \simeq \frac{1}{4}\frac{\dot{\phi}}{H} \frac{1}{M_{pl}^2}\varphi^2 
= \frac{\dot{\phi}}{H} \frac{\epsilon}{2} 
R^{(1)\,\,2} \,.
\label{relation_ord1_ord2}
\ee

The correlation function for $Q^{(2)}$ turn simplifies to 
\begin{eqnarray}
& & \langle 0 |Q^{(2)}(t, {\bf x}) Q^{(2)}(t, {\bf y})| 0 \rangle \simeq 
-\frac{\epsilon^3 M_{pl}^2}{2} \langle 0 |R^{(2)}_A(t, {\bf x}) 
R^{(2)}_A(t, {\bf y})| 0 \rangle =
\left(
\frac{\dot{\phi}}{4 H}\frac{1}{M_{pl}^2}\right)^2\frac{1}{(2 \pi)^6}
\left\{  \int
d^3 k_1 d^3 k_2 \frac{H(t_{k_1})^4}{k_1^3}\frac{H(t_{k_2})^4}{k_2^3}
 \right. \nonumber \\ 
& &\,\,\,\,\, \,\,\,\,\,\left.
\frac{1}{2 H^4}
\Theta(a H-k_1)
\Theta(a H-k_2)
\Theta(k_1-l)
\Theta(k_2-l)
\exp^{i ({\bf k_1}+{\bf k_2}) \cdot ({\bf x}-{\bf y})}+\int d^3 k_1 d^3 k_2
 \frac{H(t_{k_1})^4}{k_1^3}\frac{H(t_{k_2})^4}{k_2^3}\frac{1}{4 H^4}
\right. \nonumber\\
& &\,\,\,\,\,\,\,\,\,\, \left. \Theta(a H-k_1) \Theta(a H-k_2)
\Theta(k_1-l)
\Theta(k_2-l)
   \right\}\,.
\label{Corr_q_varphi}
\end{eqnarray}


So, on proceeding as the previous subsection, for $p \ne 0$, to the 
leading order and in an ``approximate'' fashion with the condition 
$l<<p<<a(t) H(t)$ one obtains
\be
|Q^{(2)}_p(t)|^2\simeq\left(\frac{\dot{\phi}}{4 H}\frac{1}{M_{pl}^2}\right)^2
\frac{1}{8 \pi^2} \frac{H_0^8}{H^4}\frac{1}{p^3}
\left\{ \sum_{i=0}^3 A_i \left(\ln{\frac{p}{l}}\right)^i
+g\left(\frac{l}{p}\right)+h\left(\frac{p}{a(t)H(t)}\right)
\right\} \,.
\label{q2_value}
\ee 
Substituting the coefficients $A_i$ we obtain the following result:
\begin{eqnarray}
|Q^{(2)}_p(t)|^2 &\simeq& \!\!\!\!
\left(\frac{\dot{\phi}}{4 H}\frac{1}{M_{pl}^2}\right)^2
\!\!\frac{1}{8 \pi^2} \epsilon^2
\frac{1}{p^3}
\left\{-2.8405\frac{1}{\dot{H}} H(t_p)^6+19.3975  H(t_p)^4-
14.9341 \dot{H} H(t_p)^2+ 9.7273 \dot{H}^2+4 \frac{1}{\dot{H}^2}
H(t_p)^8 \right. \nonumber \\
& & \left. \!\!\!\! 
\ln{\frac{p}{l}}+16  H(t_p)^4 \left(\ln{\frac{p}{l}}\right)^2
+\frac{16}{3}  H(t_p)^4 \left(\ln{\frac{p}{l}}\right)^3
+g\left(\frac{l}{p}\right)+h\left(\frac{p}{a(t)H(t)}\right)
\right\} \,.
\label{q2_value_developed}
\end{eqnarray}
On taking only the leading term for the coefficents of the different 
powers in $\ln(p/l)$, we obtain:
\begin{eqnarray}
|Q^{(2)}_p(t)|^2 &\simeq& 
\frac{1}{96 \pi^2} \epsilon^2  |\varphi_p(t)|^2
\left\{ 
8.5215 \, \frac{H(t_p)^2}{M_{pl}^2} + 36 \frac{H(t_p)^4}{m^2 M_{pl}^2} 
\left( \ln{\frac{p}{l}} \right) + 16 \, 
\frac{m^2}{M_{pl}^2} \left(\ln{\frac{p}{l}}\right)^2
+ \frac{16}{3} \, \frac{m^2}{M_{pl}^2} 
\left(\ln{\frac{p}{l}}\right)^3 \right\}
\,.
\label{q2_value_developed_lead}
\end{eqnarray}
Analogous considerations to those given after 
Eq.(\ref{R2A_value_developed_lead}) are also valid in this case.
Also, as we can see from Eq. (\ref{q2_value_developed}), 
one obtains a spectrum which, 
according to 
Eqs. (\ref{R2A_leading})
and (\ref{relation_ord1_ord2}), slowly 
increases in time during slow-roll as $\epsilon^3$, whereas $R^{(2)}_A$ is 
constant.
Again, we note that both the GI second order field fluctuation $Q^{(2)}$ and 
the second order curvature $R^{(2)}_A$ have a nearly scale invariant 
spectrum with a logarithmic corrections.


\section{NUMERICAL ANALYSIS}\label{two}
Let us introduce a few parameters which are useful for the numerical analysis.
The first is $N_{tot}=\ln\left(a(t_f)/a(t_i)\right)$, the total number 
of e-folds, thus we need to specify  
the time associated with the end 
of inflation. This definition may be obtained on requiring the Hubble 
parameter $H(t)$ to be equal, for example, to the inflaton mass $m$. 
This happens, in our chaotic model  
for the time $t_f$ when the slow-roll parameter $\epsilon$ is equal to $1/3$. 
We then solve the equation
\be
\epsilon(\tilde{t})=\tilde{\epsilon}
\label{EQ_general}
\ee
obtaining
\be
\tilde{t}=-\frac{H_0}{\dot{H}}-\sqrt{-\frac{1}{\tilde{\epsilon} \dot{H}}}
\ee
and the final time, defined by $\tilde{\epsilon}=1/3$, is
\be
t_f=-\frac{H_0}{\dot{H}}-\sqrt{-\frac{3}{\dot{H}}}
\ee
corresponding to a total number of e-folds given by
\be
N_{tot}=\frac{3}{2} \frac{H_0^2}{m^2}-\frac{3}{2} \,.
\ee
It is also useful to define the variable $N(t)$, which is 
 equal to the number of e-folds from
the end of inflation,
\be
N(t)=\ln\frac{a(t_f)}{a(t)}=\frac{3}{2} \frac{H(t)^2}{m^2}-\frac{3}{2}\,.
\ee
In our analysis we are interested in the fluctuations that cross the horizon 
at $N_*=55$ e-folds from the end of inflation. The time $t_*$ associated 
with $N_* \equiv \ln \left( a(t_f)/a(t_*) \right) = 55$ is:
\be
t_* =-\frac{H_0}{\dot{H}}-\sqrt{-\frac{3}{\dot{H}}-\frac{2}{\dot{H}}
N_*}
\ee
and the comoving wave-number which crosses the Hubble radius at $N_*$ is:
\be
k_{*} = a(t_*) H(t_*) = e^{-\frac{H_0^2}{2 \dot{H}}-\frac{3}{2}
-N_*}\sqrt{-3 \dot{H} -2\dot{H} N_*} \,.
\ee
If we wish to take a ``photograph'' of a fluctuation with $k_*$ on 
super-Hubble scales during inflation we should take: 
\be
\frac{1}{3}>\tilde{\epsilon}>\frac{1}{2 N_* + 3} \,.
\label{Cond_photo}
\ee

We now proceed with the numerical analysis taking $m=10^{-5} M_{pl}$. 
In particular we are interested in the comparison of second order 
fluctuations with first order ones, in order to understand for 
which condition first order
perturbation theory breaks down. We shall also plot the shape, in $k$, of 
the spectrum of fluctuations.

\subsection{Second order GI scalar field fluctuation $Q^{(2)}$}

In figure $1$ we consider,
for two different values of the initial Hubble parameter $H_0$ 
($7 m$ and $14 m$, 
respectively), the logarithm of the second order contribution to the power 
spectrum of $Q^{(2)}$:
$$
F^{(2)}(k,t)=\log_{10}{\left(\frac{1}{2 \pi^2} \frac{k^3}{M_{pl}^2} 
|Q^{(2)}_k(t)|^2\right)} \,,
$$ 
as given by 
Eq.(\ref{q2_value_developed}),
Eq.(\ref{q2_value_Green}) and the numerical solution of 
Eq.(\ref{trasf_fourier_Green_numerical}), in order 
to see the improvement of the roughest analytical approximation,
with respect to the exact numerical calculation with increasing
$H_0$, and the shape of the spectrum. 
We give the figure at an instant during inflation for which 
$\tilde{\epsilon}=1/10$.

Corresponding to those two values of $H_0$ we have the following data

\begin{itemize}

\item For $H_0=7 m$ we have $N_{\rm tot}=72$, $t_f=18/m$, 
$t_*=2.58805/m$ and $k_*/m=1.48247 \,\, 10^8$.
The time at which we ``photograph'' the
inflation is $\tilde{t}=15.5228/m$.

\item For $H_0=14 m$ we have $N_{tot}=292.5$, $t_f=39/m$, 
$t_*=23.588/m$ and $k_*/m=8.56876 \,\, 10^{103}$.
The time at which we ``photograph'' the
inflation is  $\tilde{t}=36.5228/m$.

\end{itemize}

\begin{figure}
\begin{center}
\resizebox{.8\textwidth}{!}{\includegraphics{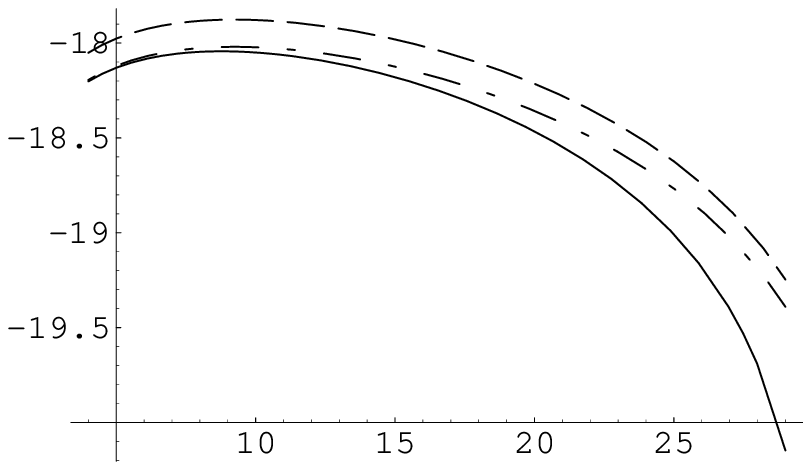}\hspace{3cm}
\includegraphics{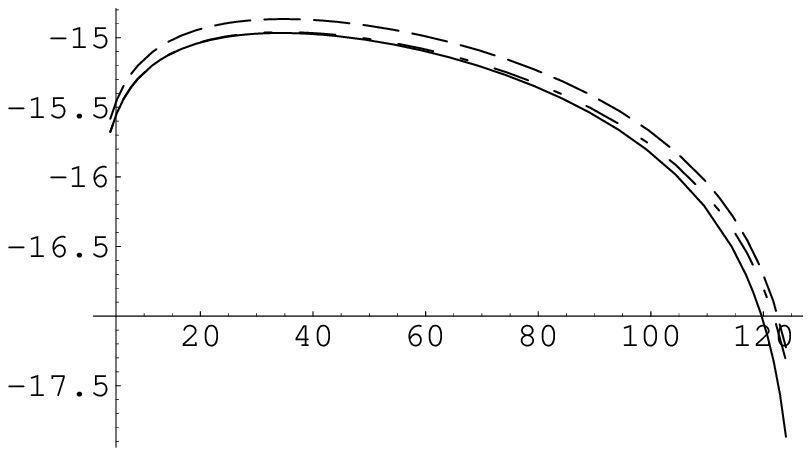}}
\rput[tl](0.2,0.9){$\log_{10}(k/m)$}
\rput[tl](-6.5,4){$F^{(2)}(k,\tilde{t})$}
\rput[tl](-8.3,0.3){$\log_{10}(k/m)$}
\rput[tl](-14.9,4.1){$F^{(2)}(k,\tilde{t})$}
\caption{We plot for $H_0=7 m$ (on the left)  and $H_0=14 m$ (on the right)
the evolution with respect to 
$\log_{10}\left(\frac{k}{m}\right)$ 
of  $F^{(2)}(k,\tilde{t})$ using
the analytic approximations Eq.(\ref{q2_value_developed})(dashed line) and 
Eq.(\ref{q2_value_Green}) (dot-dashed line) and the numerical solution of 
Eq.(\ref{trasf_fourier_Green_numerical}) (solid line).}
\end{center}
\end{figure}


In order to compare the $F^{(2)}(k,t)$ 
given by the numerical solution of 
Eq.(\ref{trasf_fourier_Green_numerical})
with the first order contribution to the power spectrum
$$
F^{(1)}(k,t)=\log_{10}{\left(\frac{1}{2 \pi^2} \frac{k^3}{M_{pl}^2} 
|\varphi_{k} (t)|^2\right)}
$$
given by Eq.(\ref{varphi_infrared}),
we give two different types of figures.
In figure $2$ we compare the 
values calculated at $k_*$ as a function of the total number of e-folds 
$N_{tot}$, at the end of inflation $t=t_f$. 
From the figure on the right (which differs from
the figure on the left only because of the bigger range of $N_{tot}$)  
we see
that second order effects are of the same order as first order effects
for $N_{tot}$ near $30000$ which corresponds to a $H_0$ near $141 m$.
Therefore for such
values first order perturbation theory breaks down 

\begin{figure}
\begin{center}
\resizebox{.8\textwidth}{!}{\includegraphics{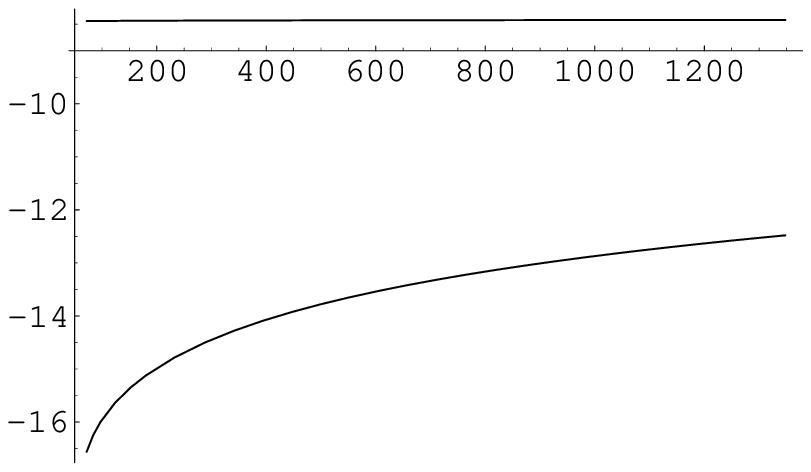}\hspace{2cm}
\includegraphics{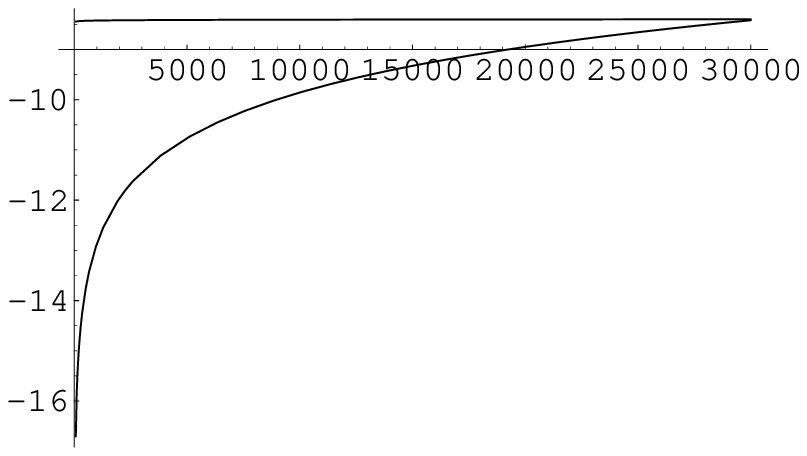}}
\rput[tl](-7.8,3.6){$N_{tot}$}
\rput[tl](-13.5,4.3){$F^{(1)}(k_*,t_{f})$}
\rput[tl](-13.5,2){$F^{(2)}(k_*,t_{f})$}
\rput[tl](0.1,3.6){$N_{tot}$}
\rput[tl](-5.5,4.3){$F^{(1)}(k_*,t_{f})$}
\rput[tl](-5.5,2){$F^{(2)}(k_*,t_{f})$}
\caption{We plot $F^{(1)}(k_*,t_f)$ and $F^{(2)}(k_*,t_f)$ with respect
the total number of e-folds $N_{tot}$.}
\end{center}
\end{figure}

In figure $3$ we show the dependence of the fluctuations on the modes, 
crossing the horizon N e-folds before the end of inflation, as seen at the
end of inflation. 
We consider three different cases corresponding to values of $H_0/m$ equal to
7, 14 and 30, compare $F^{(1)}(k_N,t_f)$ and $F^{(2)}(k_N,t_f)$, 
and vary $N$ from $55$ to $1$ with: 
\be
k_N=e^{-\frac{H_0^2}{2 \dot{H}}-\frac{3}{2}
-N}\sqrt{-3 \dot{H}
-2\dot{H} N}
\label{KNvalue}
\ee
which is the mode of the fluctuation that crosses the horizon at $N$ 
e-folds before the end of inflation.
As before the first and second order fluctuations are of the same order of 
magnitude for $H_0=141 m$.

\begin{figure}
\begin{center}
\resizebox{.4\textwidth}{!}{\includegraphics{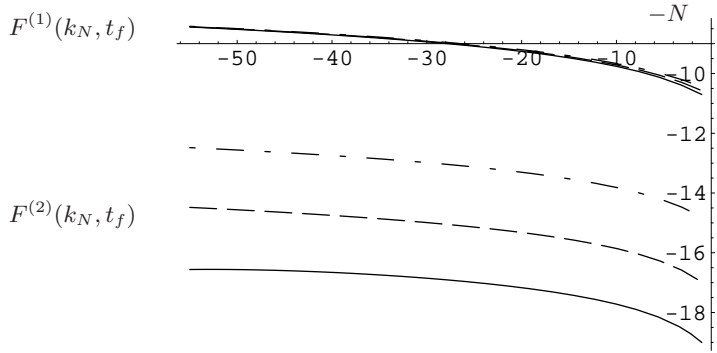}}
\rput[tl](-1,4.6){$-N$}
\rput[tl](-9.5,4.5){$F^{(1)}(k_N,t_{f})$}
\rput[tl](-9.5,2){$F^{(2)}(k_N,t_{f})$}
\caption{We plot $F^{(1)}(k_N,t_f)$ and $F^{(2)}(k_N,t_f)$ for 
$H_0=7 m$ (solid line), $H_0=14 m$ (dashed line) and $H_0=30 m$ 
(dot-dashed line) with respect $N$.}
\end{center}
\end{figure}


\subsection{Comoving curvature perturbation $R^{(2)}_A$}
Let us proceed in a analogous way for the second order curvature 
perturbation (potential) $R^{(2)}_A$.
As before in figure $4$ we shall consider two different
cases for inflation with $H_0$ equal to 7m and 14m and we shall plot 
$$
S^{(2)}(k)=\log_{10}{\left(\frac{k^3}{2 \pi^2}
|R^{(2)}_{A\,k}|^2\right)}
$$ 
the logarithm of a second order contribution to the power spectrum,  
as given by the
Eq.(\ref{R2A_value_developed}),
to exhibit  the shape of the spectrum. 
We give the picture an instant during inflation 
for which $\tilde{\epsilon}=1/10$.

\begin{figure}
\begin{center}
\resizebox{.8\textwidth}{!}{\includegraphics{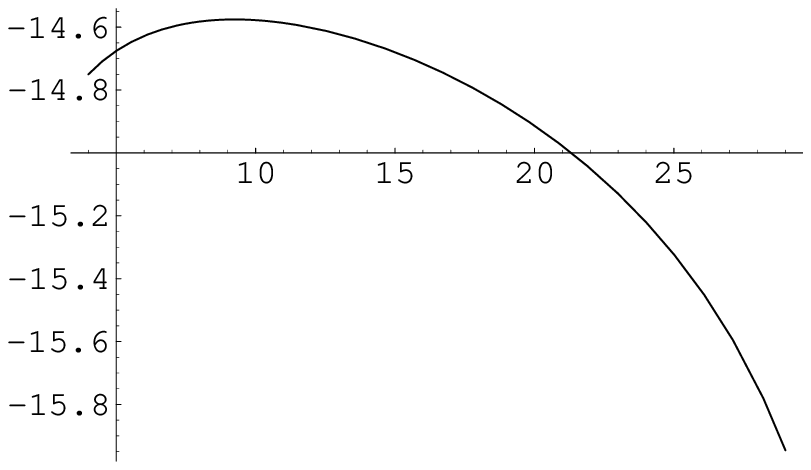}\hspace{3cm}
\includegraphics{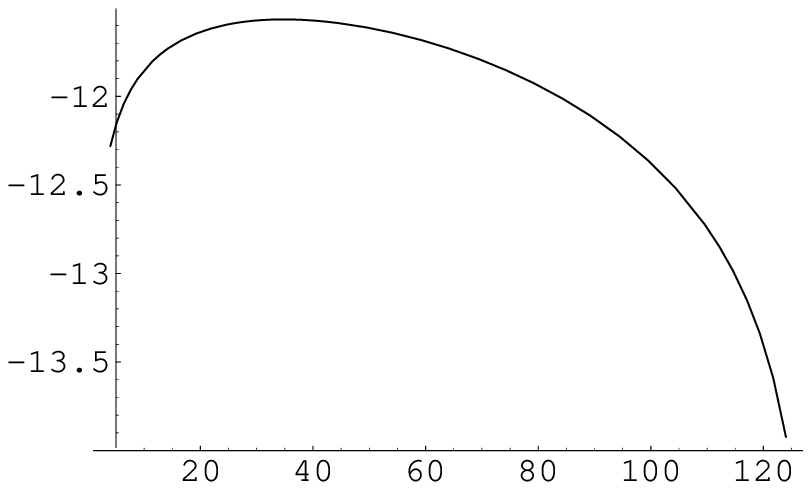}}
\rput[tl](0.2,0.3){$\log_{10}(k/m)$}
\rput[tl](-6.5,4){$S^{(2)}(k)$}
\rput[tl](-8.3,2.30){$\log_{10}(k/m)$}
\rput[tl](-14.9,4){$S^{(2)}(k)$}
\caption{We plot, for $H_0=7 m$ (on the left)  and $H_0=14 m$ (on the right),
the evolution with respect 
$\log_{10}\left(\frac{k}{m}\right)$ 
of  $S^{(2)}(k)$.}
\end{center}
\end{figure}

For the sake of comparing $S^{(2)}(k)$ 
with the first order contribution to the power spectrum 
$$
S^{(1)}(k)=\log_{10}{\left(\frac{k^3}{2 \pi^2} 
|R^{(1)}_k|^2\right)}
$$

given by Eq.(\ref{1storderspectrum}),
we give two different kinds of figures. 
In figure $5$ we compare those 
values calculated at $k_*$, as a function of the total number of 
e-folds $N_{tot}$,
at the end of inflation $t_f$. 
From the figure on the right (which again differs from
the figure on the left only because of the larger range of $N_{tot}$)  
we see
that for this case the second order effect is of the same order as the first 
order effect for $N_{tot}$ near $7560$ which corresponds to a $H_0$ near 
$71 m$. 
Therefore first order perturbation theory breaks down for a value of 
$H_0$ smaller than that obtained before and we have a stronger limit.

\begin{figure}
\begin{center}
\resizebox{.8\textwidth}{!}{\includegraphics{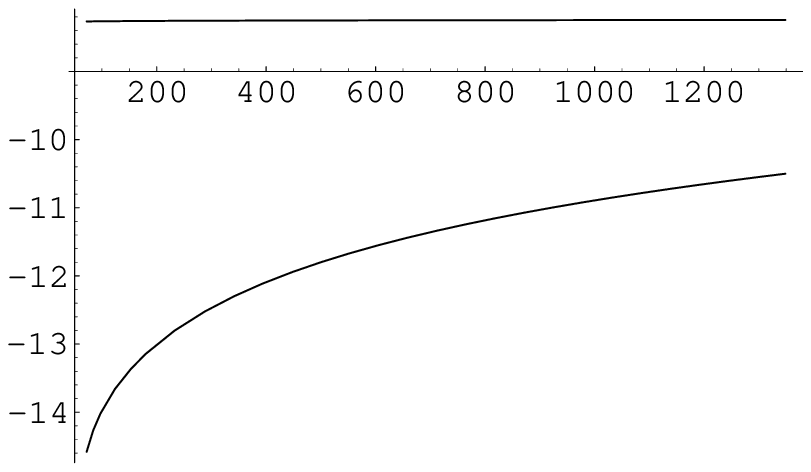}\hspace{2cm}
\includegraphics{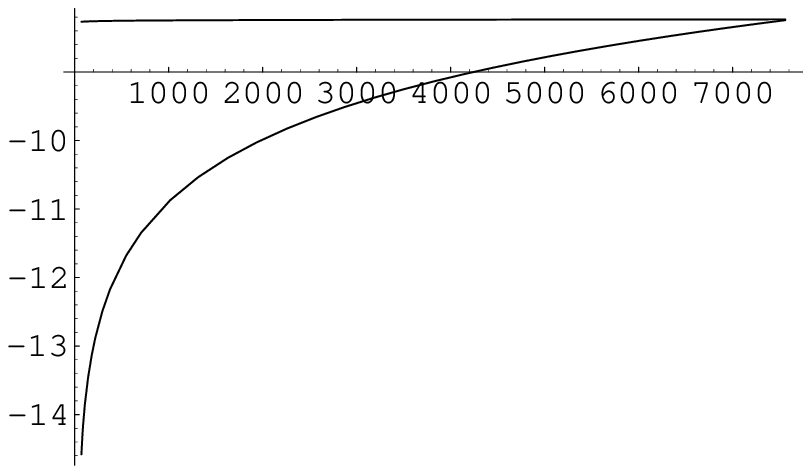}}
\rput[tl](-8,3.5){$N_{tot}$}
\rput[tl](-13.5,4.1){$S^{(1)}(k_*)$}
\rput[tl](-13.5,2.2){$S^{(2)}(k_*)$}
\rput[tl](0,3.5){$N_{tot}$}
\rput[tl](-5.5,4.1){$S^{(1)}(k_*)$}
\rput[tl](-5.2,2){$S^{(2)}(k_*)$}
\caption{We plot $S^{(1)}(k_*)$ and $S^{(2)}(k_*)$ with respect to
the total number of e-folds $N_{tot}$.}
\end{center}
\end{figure}

In figure $6$ we again consider three different value of $H_0/m$, 
namely $7$, $14$ and $30$, compare $S^{(1)}(k_N)$ and 
$S^{(2)}(k_N)$, and vary $N$ 
from $55$ to $1$ at the end of inflation. 
As before the first and second orders will be of 
the same order of 
magnitude if we take $H_0=71 m$.

\begin{figure}
\begin{center}
\resizebox{.4\textwidth}{!}{\includegraphics{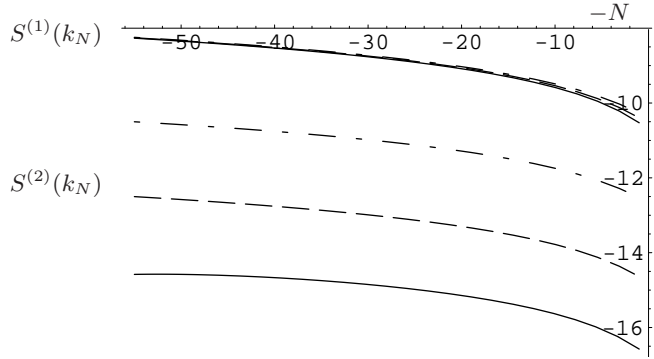}}
\rput[tl](-1,4.7){$-N$}
\rput[tl](-8.7,4.5){$S^{(1)}(k_N)$}
\rput[tl](-8.7,2.5){$S^{(2)}(k_N)$}
\caption{We plot $S^{(1)}(k_N)$ and $S^{(2)}(k_N)$ for 
$H_0=7 m$ (solid line), $H_0=14 m$ (dashed line) and $H_0=30 m$ 
(dot-dashed line) with respect to $N$.}
\end{center}
\end{figure}


Finally we wish to compare the critical initial values of $H_0$ which 
we obtain in this paper, with two different limiting values.
The first one is the value $H_{\rm br}$, obtained by requiring that 
back-reaction in the UCG gauge will become important by the end of 
inflation \cite{FMVV_2,FMVV_S}, when $H \sim m$. Such a value is:
\be
H_{br} \sim \left(16 \pi^2 \right)^{1/6} m^{2/3} 
M_{pl}^{1/3} \,.
\label{Hbr}
\ee
The second one is the value obtained by the self-reproduction argument 
\cite{self}
\be
H_{sr} \simeq (8 \pi)^{3/4}\left(\frac{1}{6}\right)^{1/2} m^{1/2} 
M_{pl}^{1/2} \,.
\label{Hsr}
\ee
This comparison is given in Table I.

\begin{table}[htbp]
\begin{center}
\begin{tabular}{|c|l|}
\hline
$H_0$ limit for $R^{(2)}_A$ & 71 m \\
\hline
$H_0$ limit for $Q^{(2)}$ & 141 m \\
\hline
$H_{\rm br}$ & 108 m \\
\hline
$H_{\rm sr}$ & 1449 m \\
\hline
\end{tabular}
\end{center}
\caption{Comparison of the values of $H_0$ for which first order 
perturbation theory break down with $H_{\rm br}$ and $H_{\rm sr}$.}
\end{table}

We stress that neither of the two values of $H_0$ or of $H_{\rm br}$ are 
related to the self-reproduction scale \cite{self}. The values of $H_0$ 
obtained by considering the breakdown of linear perturbation theory 
in this paper are of the same order of magnitude 
as $H_{\rm br}$ (see also \cite{ABM} for an analogous value). 
This fact is not  surprising: both these calculations include second-order 
cosmological perturbations. 
The results for back-reaction were considered ambiguous since 
it is not simple to demonstrate the slow-down of the inflationary expansion 
in terms of a GI quantity.
On the other hand in the present paper, the breakdown of first order 
perturbation theory is found for GI variables 
in the long-wavelength limit. 

The results summarized in the Table I and derived from figures 2, 3, 5 and 6
can also be obtained directly from
Eq.(\ref{R2A_value_developed_lead}) and Eq.(\ref{q2_value_developed_lead}).
In those cases and for $H_0>>m$ on using Eq.(\ref{KNvalue}) 
the leading logarithm values (that is the term with
$\left(\ln{\frac{p}{l}}\right)^3$) of
$|R^{(2)}_{A\,p_N}|^2$ and $|Q^{(2)}_{p_N}(t_f)|^2$ are given by:
\begin{eqnarray}
|R^{(2)}_{A\,p_N}|^2 &\simeq& 
\frac{3}{4 \pi^2} |R^{(1)}_{p_N}|^2 \frac{H_0^6}{M_{pl}^2  m^4}
\label{R2A_value_developed_leadfinale}
\\
|Q^{(2)}_{p_N}(t_f)|^2 &\simeq& 
\frac{3}{16 \pi^2} \epsilon(t_f)^2 |\varphi_{p_N}(t_f)|^2 
\frac{H_0^6}{M_{pl}^2  m^4}
\label{q2_value_developed_leadfinale}
\end{eqnarray}
and the first and second orders are comparable for
\be
H_0 \simeq m \left( \frac{2 \pi}{\sqrt{3}} \frac{M_{pl}}{m}\right)^{1/3}
\label{H0R}
\ee
and 
\be
H_0 \simeq m \left( \frac{12 \pi}{\sqrt{3}} 
\frac{M_{pl}}{m}\right)^{1/3}
\label{H0Q}
\ee
respectively,
where Eq.(\ref{H0R}) nearly give the same value as Table I 
(namely near $71 m$), while Eq.(\ref{H0Q}) only gives the same order of
magnitude (actually $130 m$), which is not surprising considering the 
approximations made.


\section{Conclusions}

We have studied second order cosmological perturbations for a chaotic 
$\frac{m^2}{2} \phi^2$ inflationary model, and considered three second 
order GI measures of curvature perturbations. 

For the GI curvature perturbation studied and for the second order GI 
scalar field flcutuation presented in \cite{Malik}, 
we have found, for all scales of interest (see, for example, 
Eq.(\ref{R2A_value_developed_leadfinale}) and 
Eq.(\ref{q2_value_developed_leadfinale})),
that the amplitude of the spectrum of quadratic 
corrections grows with the total number of e-folds $N_{\rm tot}$. 
For $Q^{(2)}$ we also obtain a mild time dependence associated with the 
growth of the slow-roll parameter $\epsilon$. Since both the quantities 
studied here are GI, this dependence on $N_{\rm tot}$ cannot be a gauge 
artifact. 

The spectrum of those second order GI variables obtained by
Fourier trasforming the curvature-curvature quantum correlation function is 
nearly scale-invariant, with additional logarithmic corrections with respect 
to the first order spectrum. On comparing the first and 
the second order contributions for
the two different cases considered, one finds two different values for the 
initial Hubble parameter $H_0$ beyond which first order perturbation 
theory breaks
down. 
We found limits which are of the same order of magnitude as that one 
found by back-reaction in the UCG \cite{FMVV_2}. 
Neither the back-reaction limit nor those found here have anything to do 
with the self-reproduction scale.
 
In the curvature-curvature quantum correlation function,
the cross terms involving first and third order perturbation may also
contribute. The third order curvature perturbation is clearly beyond 
the scope of 
the present paper and will be the subject of future work. 
Unless such a term  cancels or dominates the growth we have found, our results 
imply that in single field models the 
amplitude of the spectrum of intrinsic non-gaussianities generated during 
the slow-roll stage will depend on the total number of e-folds which inflation 
has lasted.
\vspace{6mm}

\hspace{6cm}{\bf ACKNOWLEDGEMENTS}

\vspace{6mm}

We wish to thank R. Abramo, S. Matarrese, A. A. Starobinsky for discussions
and K. A. Malik for  useful correspondence.


\section{Appendix A: Gauge Transformations}\label{App_A}

The gauge in our formulation (see Eq.(\ref{metric_second})) is not fixed, 
since one
can eliminate two scalars among the metric and the inflaton degrees of
freedom. After a general analysis we shall be interested in making a
gauge choice in order to match previously studied situations.
An infinitesimal coordinate trasformation to second order is \cite{bruni}:
\be
x^\mu \rightarrow \tilde{x}^\mu= x^\mu + \epsilon^\mu_{(1)} +\frac{1}{2}
\left(\epsilon^{\mu}_{(1),\nu}
\epsilon^{\nu}_{(1)} + \epsilon^{\mu}_{(2)}\right) \,,
\label{Trasf_gauge_coord}
\ee
where $\epsilon_{(1)}$ and $\epsilon_{(2)}$ are the coordinate changes to
first an second order, respectively, and $\epsilon_{(2)}^i = \partial^i
 \epsilon^s_{(2)}$ and
$\epsilon_{(1)}^i =\partial^i \epsilon^s_{(1)}$.
It induces the following change in a geometric object
 $T = T^{(0)} + T^{(1)}
+ T^{(2)}$ :
\be
T^{(1)} \rightarrow \tilde{T}^{(1)}=
T^{(1)} - \mathcal{L}_{\epsilon_{(1)}} T^{(0)}
\label{Trasf_tens_ord1}
\ee
\be
T^{(2)} \rightarrow  \tilde{T}^{(2)}=
T^{(2)} - \mathcal{L}_{\epsilon_{(1)}} T^{(1)} +
\frac{1}{2}
\left(\mathcal{L}_{\epsilon_{(1)}}^2 T_0 -
\mathcal{L}_{\epsilon_{(2)}} T_0 \right)
\label{Trasf_tens_ord2}
\ee
which leads to the following general gauge trasformation for our scalar
perturbations. To first order we have:
\be
\tilde{\varphi}= \varphi - \epsilon^0_{(1)} \dot{\phi}
\label{Trasf_inf_ord_uno}
\ee
\be
\alphat = \alpha - \dot \epsilon^0_{(1)}
\ee
\be
\betat = \beta - \frac{2}{a} \epsilon^0_{(1)} +2 a \dot \epsilon^s_{(1)}
\ee
\be
\tilde{\psi} =\psi+ H \epsilon^0_{(1)}+\frac{1}{3} 
\nabla^2\epsilon^s_{(1)}
\ee
\be
\tilde{E}=E -2\epsilon^s_{(1)} \,.
\label{E_transformation}
\ee
To second order one finds:
\be
\tilde{\varphi}^{(2)}=
\varphi^{(2)} - \epsilon^0_{(1)} \dot{\varphi} + \frac{1}{2}
\left[\epsilon^0_{(1)}
\left(\epsilon^0_{(1)}\dot{\phi}\right)^.
- \epsilon^0_{(2)} \dot{\phi}\right] -\partial_k \varphi \partial^k\epsilon^s_{(1)} \,,
\label{Trasf_inf_ord_due}
\ee
\be
\alphat^{(2)} = \alpha^{(2)}- \epsilon^0_{(1)} \dot{\alpha}-2 \alpha
\dot \epsilon^0_{(1)}-\frac{1}{2}\dot \epsilon^0_{(2)}+\frac{1}{2}
\epsilon^0_{(1)}\ddot \epsilon^0_{(1)} +\left(\dot 
\epsilon^0_{(1)}\right)^2
\ee
\be
\tilde{\beta}^{(2)}= \beta^{(2)} -\frac{1}{a} \epsilon_{(2)}^0
+ a \dot{\epsilon}^s_{(2)} +\frac{1}{2a}\frac{d}{d t} (\epsilon_{(1)}^0)^2
+\frac{\partial^i}{\nabla^2}
\left[ \frac{2}{a} \partial_i\epsilon_{(1)}^0 \dot{\epsilon}_{(1)}^0
- \epsilon_{(1)}^0  \partial_i \dot{\beta}-
H \epsilon_{(1)}^0 \partial_i \beta - \dot{\epsilon}_{(1)}^0 \partial_i
\beta - \frac{4}{a} \alpha \partial_i \epsilon_{(1)}^0
\right]
\ee
\begin{eqnarray}
\tilde{\psi}^{(2)} &=& \psi^{(2)} - \epsilon_{(1)}^0 \left(2 H \psi+\dot{\psi}
\right) +
\frac{H}{2}\epsilon^0_{(2)} -
\frac{H}{2} \epsilon^0_{(1)} \dot \epsilon^0_{(1)} -
\frac{\epsilon^{0 \, 2}_{(1)}}{2} \left( \dot H + 2 H^2 \right) +
\frac{1}{6a^2} \partial^i \epsilon^0_{(1)} \, \partial_i \epsilon^0_{(1)}
-\frac{1}{6a} \partial^i \beta \partial_i \epsilon^0_{(1)}
+\frac{1}{6} \nabla^2 \epsilon^s_{(2)}- \nonumber \\
& & -\frac{1}{12} \nabla^2 \left( \partial_k \epsilon^s_{(1)} \partial^k \epsilon^s_{(1)}
\right) -\frac{2}{3} H \epsilon^0_{(1)} \nabla^2 \epsilon^s_{(1)}-\frac{1}{6}
 \epsilon^0_{(1)} \nabla^2 \dot \epsilon^s_{(1)}-\frac{1}{2} H   \partial^k
\epsilon^s_{(1)} \partial_k \epsilon^0_{(1)} -\frac{1}{6} \partial^k
\dot \epsilon^s_{(1)} \partial_k \epsilon^0_{(1)}- \nonumber \\
& & -\partial^k \epsilon^s_{(1)} \partial_k \psi - \frac{2}{3} \psi \nabla^2 \epsilon^s_{(1)}
+\frac{1}{3}  D_{kj} E \partial^j \partial^k \epsilon^s_{(1)}
\end{eqnarray}
\begin{eqnarray}
\tilde{E}^{(2)} &=& E^{(2)}+\frac{1}{2a} \beta \epsilon^0_{(1)} -
\frac{1}{2a^2} (\epsilon^0_{(1)})^2 - \epsilon^s_{(2)}+
\frac{1}{2} \partial^k \epsilon^s_{(1)} \partial_k \epsilon^s_{(1)}+
\frac{1}{2} \dot \epsilon^s_{(1)} \epsilon^0_{(1)} -\partial_k E
 \partial^k \epsilon^s_{(1)}+
\nonumber \\
& &  +\frac{3}{2}\frac{\partial^i \partial^j}{(\nabla^2)^2}\left[-2 H
\epsilon^0_{(1)} D_{ij} E -\epsilon^0_{(1)} D_{ij} \dot{E}
+ \frac{1}{a^2}\epsilon^0_{(1)} D_{ij} \epsilon^0_{(1)} -
 \frac{1}{2a} \left(
\epsilon^0_{(1)} D_{ij} \beta + \beta  D_{ij} \epsilon^0_{(1)}
\right)
+4 H \epsilon^0_{(1)} D_{ij} \epsilon^s_{(1)} + \right. \nonumber \\
& & \left.  +
\frac{1}{2} \left( \epsilon^0_{(1)} D_{ij} \dot \epsilon^s_{(1)} - \dot \epsilon^s_{(1)}
 D_{ij} \epsilon^0_{(1)} \right) + 4 \psi D_{ij} \epsilon^s_{(1)} +
 \partial_k E  D_{ij} \partial^k \epsilon^s_{(1)} +\frac{2}{3}
\nabla^2  E  D_{ij} \epsilon^s_{(1)}
\right] \, .
\label{E2_transformation}
\end{eqnarray}


\section{Appendix B: $A_i$ coefficient}\label{App_B}

The coefficients $A_i$ are given by:
\begin{eqnarray}
A_0(k)&=& 16 \, \epsilon_0^2 \, (\zeta (3)-1)C_1(k)-10.198 \, 
\epsilon_0^2 \,C_2(k)
+24.656 \, \epsilon_0^2 \, C_3(k)+ \epsilon_0 \left(8-\frac{2}{3}\pi^2\right)
C_1(k) C_4(k)
\nonumber \\
&& -9.30635 \, \epsilon_0 \, C_2(k) C_4(k) + 33.333 \, \epsilon_0 \, 
C_3(k) C_4(k) +\left(\frac{\pi^2}{3}-4\right)  C_2(k) C_4(k)^2+6.07344 \, 
 C_3(k) C_4(k)^2
\nonumber \\
A_1(k)&=& 4  \, C_1(k) C_4(k)^2-8 \, C_2(k) C_4(k)^2
\nonumber \\
A_2(k)&=&  4\, \epsilon_0 \, C_1(k) C_4(k)- 8 \, \epsilon_0 \, C_2(k) C_4(k)+2
 \, C_2(k) C_4(k)^2+8  \,C_3(k) C_4(k)^2
\nonumber \\
A_3(k)&=& \frac{8}{3} \, \epsilon_0^2 \,C_1(k)-\frac{16}{3} \, \epsilon_0^2 \,
C_2(k) +\frac{8}{3} \,C_3(k) \,C_4(k)^2
\label{A_i_coefficient}
\end{eqnarray}
where
\begin{eqnarray}
C_1(k)&=& \frac{H(t_k)^4}{H_0^4}-4  \epsilon_0\frac{H(t_k)^2}{H_0^2}
+8 \epsilon_0^2
\nonumber \\
C_2(k)&=&  -2 \epsilon_0\left(\frac{H(t_k)^2}{H_0^2}-2 \epsilon_0\right)
\nonumber \\
C_3(k)&=&  \epsilon_0^2
\nonumber \\
C_4(k)&=& \frac{H(t_k)^2}{H_0^2}
\label{C_J_coefficient}
\end{eqnarray}
so one obtains

\begin{eqnarray}
A_0(k)&=& \left(16-\frac{4}{3}\pi^2\right)\epsilon_0 \frac{H(t_k)^6}{H_0^6}+
\left(4 \pi^2+16 \, \zeta (3)-39.3139\right)\epsilon_0^2 \frac{H(t_k)^4}{H_0^4}
+\left(144.5036-\frac{16}{3} \pi^2-64 \, \zeta (3)\right) 
\nonumber \\
&& 
\epsilon_0^3
\frac{H(t_k)^2}{H_0^2}+
\left(128 \, \zeta (3)-144.136\right) \epsilon_0^4
\nonumber \\
A_1(k)&=& 4 \frac{H(t_k)^8}{H_0^8}
\nonumber \\
A_2(k)&=& 16 \epsilon_0^2 \frac{H(t_k)^4}{H_0^4}
\nonumber \\
A_3(k)&=& \frac{16}{3} \epsilon_0^2 \frac{H(t_k)^4}{H_0^4}
\label{A_i_final_coefficient}
\end{eqnarray}

with the relation $3 \epsilon_0 < \frac{H(t_k)^2}{H_0^2} < 1$, which 
is valid during the inflationary era (which, for us, ends 
at $\epsilon(t)=1/3$).


\end{document}